# Twofold exp and log

Evgeny Latkin
2015 Feb 17

Project site: https://sites.google.com/site/yevgenylatkin/
Author email: yevgeny.latkin@gmail.com

**Abstract**: This article is about twofold arithmetic [1, 2]. Here I introduce algorithms and experimental code for twofold variant of C/C++ standard functions exp() and log(), and expm1() and log1p(). Twofold function $y_0 + y_1 \approx \mathrm{f}(x_0 + x_1)$ is nearly 2x-precise so can assess accuracy of standard one. Performance allows assessing on-fly: twofold texp() over `double` is ~10x times faster than expq() by GNU quadmath.

**Contents**

# Overview

Software industry needs to mitigate cost of programming as computers penetrate everywhere involving mass of coders. Managed runtimes allow coding easier and faster mitigating risks of errors, and frankly, somewhat soften qualification requirements. However, programming math still requires too high skills.

Intervals may look obvious technique for controlling rounding errors, but actually are too strict for that. Intervals prove correctness of solution, while typical numeric computing methods are formally incorrect due to rounding errors. In other words, intervals are not directly applicable to habitual computer math.

Twofolds approach is compromise. Twofolds assess deviation between exact and approximate solutions, but cannot guarantee it. We accept the risk that twofold estimate may occur completely wrong. The lack of guarantees is the cost we pay for addressing all math methods including formally incorrect.

Twofolds technique is very close to well-known double-double arithmetic and similar calculi [4-6]; and reuses Dekker formulas [7], probably fastest though not most accurate among such algorithms.

Given a floating-point format, like `float` or `double` of C/C++, and a real value $x$, a twofold represents $x$ with formal sum of floating-point numbers $x_0 + x_1$ so that $x_0 \approx x$ is possibly nearest, correctly rounded of $x$ in ideal case, and $x_1$ assesses deviation $\Delta x_0 = x - x_0$.

Twofolds arithmetic as proposed in [1-2] defines operations like sum $z_0 + z_1 = (x_0 + x_1) \oplus (y_0 + y_1)$, so that $z_0 = x_0 \oplus y_0$ equals to ordinary floating-point sum, and $z_1$ assesses its deviation $\Delta z_0 = z - z_0$ from exact sum $z = (x_0 + x_1) + (y_0 + y_1)$. Effectively, twofold $z_0 + z_1$ approximates $z$ with nearly 2x-higher precision if deviation $\Delta z_0$ is small comparing $z_0$.

A twofold function $z_0 + z_1 \approx \mathrm{f}(x_0 + x_1)$ main part $z_0 \approx \mathrm{f}(x_0)$ reproduces standard function from C/C++ library, and $z_1$ assesses deviation $\Delta z_0 = z - z_0$ from exact result $z = \mathrm{f}(x)$ for $x = x_0 + x_1$. We need $z_0$ be equal to standard result on bitwise basis; assuming standard library itself reproduces results bitwise.

Idea is assessing accumulation of inaccuracy in a long chain of floating-point calculations, by effectively recalculating everything with nearly 2x-higher precision and tracking the deviation. Performance is very important if we want to track in on-fly manner, in parallel with main computations.

Here I'd like to propose fast algorithms and experimental C/C++ implementation of twofold variant of standard functions exp() and log(), and accompanying expm1() and log1p(). My goal is performance; twofold functions over `double` should work ~10x times faster than GNU quadmath library on x86-64.

In addition, I construct twofold functions over `float` data. This cannot have practical sense, as obviously we can assess accuracy of single-precision function with its double-precision variant from standard math library. However, it still seems interesting for me to investigate twofold functions over `float` as well.

Twofold exponent section below explains reducing evaluation of twofold exp() and expm1() to simpler functions pexp0() and pexpm10(), which supply $e^x$ and $e^x - 1$ with nearly 2x-higher precision provided argument $x$ is of 1x-precision. These functions allows fast computing basing on Taylor series.

Twofold logarithm section explains computing twofold log() and log1p() by inverting exp() and expm1() with Newton method. Using 1x-precise logarithms from standard library as initial approximation allows fast inverting, normally with just one iteration.

Experimental code section explains C/C++ interface and experimental implementation. Analysis section observes performance and accuracy testing results. Conclusion positions twofolds as managed runtime for productive programming of math.

Twofolds update section is about changes in twofolds basics, including important bug fixed in "Twofold arithmetic" [1] formulas, and corresponding bugs in software. Another major update is supporting older processors not supporting fast fused-multiply-add (FMA) in hardware.

How to download section explains twofolds project Web site; downloading is free for non-commercial and academic use. Please do not hesitate to contact me if any proposals and/or questions. Here is the link to project Web site and my email. See also links [1-3] in the References section:

Project site: https://sites.google.com/site/yevgenylatkin/
Email: yevgeny.latkin@gmail.com

# Twofold exponent

## Method summary

Given a floating-point format, presumably C/C++ `double`, let $x = x_0 + x_1$ be a twofold of this format. Define twofold functions $z_0 + z_1 = \mathrm{texp}(x_0 + x_1)$ like approximation of exact $e^x$ such that $z_0$ bitwise equals to $\exp(x_0)$ computed with C/C++ standard library and $z_1$ assesses deviation $\Delta z_0 = e^x - z_0$.

Ideally, $z_1$ would be the correctly rounded to nearest of $\Delta z_0$, but we do not require such strictness. As well, we do not require $z_0$ be correctly rounded of exact $e^x$. Math library may allow result be incorrect sometimes, typically by not more than 1/2 to 1 of ULP (unit in last position) of result.

Such definition implicitly assumes bitwise reproducibility of library results. This assumption is too tight, as different versions of same math library may violate it. Realistic assumption is that $\exp(x_0)$ is bitwise reproducible with same version and build (compiled binaries) of math library linked to your program.

We would require a bit less than that. Let us assume that math library reproduces results during one run of a program. Particularly, this allows another run to link with different version of dynamic math library.

Let us define twofold function $z_0 + z_1 = \mathrm{texpm1}(x_0 + x_1)$ similarly for $e^x - 1$ and library $\mathrm{expm1}(x_0)$.

Computing $z_0$ is easy, just call library function, $z_0 = \exp(x_0)$ or $z_0 = \mathrm{expm1}(x_0)$. So let us compute $z_1$. Note that we need $z_1$ with standard 1x-precision.

For `texp`(), we have $z_1 \approx \Delta z_0 = e^x - z_0 = e^{x_0+x_1} - z_0 = (e^{x_0+x_1} - e^{x_0}) + (e^{x_0} - z_0) = u + v$. Because $e^{x_0+x_1} - e^{x_0} = e^{x_0}(e^{x_1} - 1)$, product $\exp(x_0) \times \mathrm{expm1}(x_1)$ computed via standard library assesses $u$ with 1x-precision. So we could easily compute $z_1$ like $u + v$ if we knew $v = e^{x_0} - \exp(x_0)$, which we would know if we knew $e^{x_0}$ with 2x-precision.

For texpm1(), similarly $z_1 \approx \Delta z_0 = (e^{x_0+x_1} - 1) - z_0 = (e^{x_0+x_1} - e^{x_0}) + \left((e^{x_0} - 1) - z_0\right) = u + w$. Here $u$ is same as above and can be computed with 1x-precision like $\exp(x_0)\,\mathrm{expm1}(x_1)$, and we can deduce $w$ if we knew $e^{x_0} - 1$ with 2x-precision.

Let us define auxiliary function $v_0 + v_1 = \mathrm{pexp0}(x_0)$, which accept 1x-precise argument $x_0$ and returns coupled-precision approximation for $e^{x_0}$, so that $v_0 \approx e^{x_0}$ and $v_1 \approx \Delta v_0 = e^{x_0} - v_0$. We do not expect $v_0$ to equal $\exp(x_0)$; and assess $v = e^{x_0} - \exp(x_0)$ like $v_1 + (v_0 - \exp(x_0))$ with 1x-precision.

Similarly, define $w_0 + w_1 = \mathrm{pexpm10}(x_0)$, and assess $w = (e^{x_0} - 1) - z_0$ as $w_1 + (w_0 - \mathrm{expm1}(x_0))$.

Next major idea is computing of `pexp0`() and `pexpm10`() via combining table lookup and Taylor series.

Reason for preferring Taylor polynomials over Chebyshev or minimax is performance. Despite of higher polynomial degree $N$, with Taylor method we can make its coefficients 1x-precise via norming by $N!$, like $e^y N! \approx N! + N!\,y + N!\,y^2/2 + \cdots + y^N$, so make 2x-precise computations with twofolds much faster.

Consider Horner scheme, $e^y N! \approx \left(\ldots\left((y + N)y + N(N-1)\right)y + \cdots\right)y + N!$, assuming $N \ldots (N - n)$ and $y$ exactly representable as 1x-precisision floating-point numbers. Recalling twofold fast arithmetic formulas from [1], let us estimate cost of computing this scheme in twofolds.

Twofold sum of $y + N$ would cost only 3 of basic add/subtract operations, because $N \geq |y|$ for small $y$. Multiplying twofold partial sum by 1x-precise $y$ would cost 2 multiplications, 1 FMA, and 1 summation. Further summation with 1x-precise coefficient $N(N-1)$ would cost 7 of basic operations. And so on.

In overall, such Horner scheme of degree $N > 1$ would cost $2N$ multiplications, $N$ of FMA, and $8N + 3$ of add/subtract operations, so $11N + 3$ operations totally. However, modern processors can multiply in parallel with adding/subtracting; so critical path is $8N + 3$ of add/subtract operations.

For degree $N = 12$, critical path is 99 operations. Argument reduction and result reconstruction would increase this cost to around 120 operations in overall. So without vectoring for SIMD, performance must be around 20 million function calls per second per CPU core on a 2.5 GHz processor, like my laptop. This must allow $\text{texp}(x_0 + x_1)$ operate at ~10 millions per second, ~10x times faster than GNU quadmath.

My experimental code confirms these estimates: $\text{pexp0}(x_0)$ of `double` shows ~18 millions-per-second and $\text{texp}(x_0 + x_1)$ shows ~12 millions, so outperforms quad-precision $\text{expq}(x)$ by at least 15x times. (GNU quadmath shows ~0.8 million function calls per second in exponent and logarithm on my laptop.)

Actually, Taylor series is faster than theoretical estimate, as we can compute a few of highest-degree terms of Taylor polynomial with 1x-precision. But argument reduction takes more than I expected.

Another reason to prefer Taylor series is, that summing/multiplying of twofold by 1x-precise is stricter than twofold-by-twofold operation. Thus, evaluating via Taylor polynomial must be more accurate.

According to my testing with ~1 million random samples, average inaccuracy of $z_0 + z_1 = \text{texp}(x_0 + x_1)$ over `double` type appears within $2^{-100}$ (100+ significant bits) if result is not subnormal, and maximal inaccuracy fits $2^{-95}$ (95+ bits). Such average allows 47+ bits for $z_1$, which must be enough for twofolds.

## Function pexp0()

Consider in details computing twofold $\text{pexp0}(x)$ for 1x-precise argument $x$ of `double` or `float` type.

Let result be 0 or infinite, if $x$ is too small or too large, such that $e^x$ cannot fit the floating-point format. For `double` type, lower boundary is if $x < \ln 2^{-1074} \approx -744.44$, and upper is $x \geq \ln 2^{1024} \approx 709.78$. For `float`, lower boundary is $\ln 2^{-149} \approx -103.28$ and upper is $\ln 2^{128} \approx 88.72$.

Suppose $x$ is within boundaries. Let us decompose it like $x = 2^L m + 2^{-K} n + y$, where positive integers $K$ and $L$ are parameters of the method, $m$ and $n$ are integers of same sign that $x$, and floating-point $y$ is of same or opposite sign and $|y| < 2^{-K}/2$. Important, that we can compute such $y$ exactly.

<u>Algorithm</u>: Decomposing $x = 2^L m + 2^{-K} n + y$
(1) Integer $M = \text{round}(x \cdot 2^K)$ -- round to nearest
(2) Floating $y = x - M/2^K$ -- exactly (by Sterbenz lemma)
(3) Integer $m = \lfloor |M|/2^{L+K} \rfloor \cdot \text{sign}(M)$ -- upper bits of $|M|$
(4) Integer $n = |M| (\text{mod } 2^{L+K}) \cdot \text{sign}(M)$ -- lower bits of $|M|$

Idea is computing $\text{pexp0}(x)$ like product of three parts, $E \approx e^{2^L m}$, $C \approx e^{n/2^K}/N!$, and $T \approx e^y N!$, each represented as twofold and calculated with 2x-precision. Get $E$ and $C$ from precomputed tables, and compute $T$ with Taylor polynomial of degree $N$ like explained above.

For `double` type, method parameters might be $L = 2$ and $K = 5$, so that $|y| < 1/64$. Degree $N = 12$ is enough for such $y$ to guarantee Taylor polynomial is accurate to $3.85 \cdot 10^{-34}$. Note that twofold cannot be more accurate than $2^{-106} \approx 1.23 \cdot 10^{-32}$ if basic type is standard IEEE-754-2008 binary64 (`double`).

Such parameters assume table for coefficients $E$ of 364 entries for $-744 \leq 4m \leq 708$ and, and table for $C$ of 257 entries for $-128 \leq n \leq 128$. Those tables overall size would be around 10K bytes.

If basic type is `float`, parameters might be $L = 1$ and $K = 5$, so that $|y| < 1/64$. Degree $N = 6$ would be enough for Taylor series be accurate like $4.52 \cdot 10^{-17}$. Note that twofold of binary32 (`float`) cannot be more accurate than $2^{-48} \approx 3.55 \cdot 10^{-15}$.

Table for $E$ would need 96 entries for $-102 \leq 2m \leq 88$, and table for $C$ would contain 129 entries for $-64 \leq n \leq 64$. Such tables would take around 2K bytes.

Balance of table sizes and polynomial degree may be subject for further optimization.

Note that this algorithm cannot ensure bitwise reproduction of $\exp(x_0)$ from standard math library.

## Function pexpm10()

If argument $|x| \geq \ln 2$, we calculate $\mathrm{pexpm10}(x)$ just like $\mathrm{pexp0}(x) - 1$. For $|x| < \ln 2$, combine table lookup with Taylor series similarly to above. Decompose exactly $x = 2^{-K}n + y$, here $|y| < 2^{-K}/2$.

> Algorithm: Decomposing $x = 2^{-K}n + y$
> (1) Integer $n = \mathrm{round}(x \cdot 2^K)$ -- round to nearest
> (2) Floating $y = x - n/2^K$ -- exactly (by Sterbenz lemma)

Quickly compute $T \approx e^y - 1$ with Taylor series, by Horner scheme similarly to above except do not add final term $N!$, and multiply result by $1/N!$ as we cannot move this normalizing coefficient under term $C$.

Tabulate $C \approx e^{n/2^K} - 1$, and compute $\mathrm{pexpm1d}(x)$ with following formula, where $c = n/2^K$:

$$e^{c+y} - 1 = (e^c - 1)(e^y - 1) + (e^c - 1) + (e^y - 1)$$

Method parameter might be $K = 7$, so that $|y| < 1/256$. Taylor polynomial degree $N = 10$ enough for `double`, and $N = 5$ for `float`. Table for $C$ would include 179 entries for $-\ln 2 \leq n/2^K \leq +\ln 2$, which would take around 2.8K bytes for `double` and 1.4K bytes for `float`.

Note that corner elements of table for $C$ would be a little bit outside interval $[-\ln 2, +\ln 2]$, that is for maximal $n = 89$, value of $n/2^K$ a little bit exceeds $\ln 2$.

Balance of table size and polynomial degree may be subject for further optimization.

This algorithm cannot ensure bitwise reproducing $\mathrm{expm1}(x_0)$ from standard library.

## Algorithm summary

In this subsection, I enlist all functions for twofold exponentiation and write-down algorithms explicitly, for ease of referencing. Here is the list:

| Function | Description |
|---|---|
| $z_0 + z_1 = \mathrm{pexp0}(x_0)$ | Coupled exponent of "dotted" argument $x_0$, assume $x_1 = 0$ |
| $z_0 + z_1 = \mathrm{texp0}(x)$ | Twofold exponent of "dotted" $x_0$, ensure $z_0 = \exp(x_0)$ |
| $z_0 + z_1 = \mathrm{texp}(x_0 + x_1)$ | Twofold exponent of twofold argument $x_0 + x_1$ |
| $z_0 + z_1 = \mathrm{texpp}(x_0 + x_1)$ | Twofold exponent of coupled argument $x_0 + x_1$ |
| $z_0 + z_1 = \mathrm{pexp}(x_0 + x_1)$ | Coupled exponent of coupled 2x-precise argument $x_0 + x_1$ |
| $z_0 + z_1 = \mathrm{pexpm10}(x)$ | Twofold $e^x - 1$ of "doted" argument $x_0$, assume $x_1 = 0$ |
| $z_0 + z_1 = \mathrm{texpm10}(x)$ | Twofold $e^x - 1$ of "doted" $x_0$, ensure $z_0 = \mathrm{expm1}(x_0)$ |
| $z_0 + z_1 = \mathrm{texpm1}(x_0 + x_1)$ | Twofold $e^x - 1$ of twofold argument $x_0 + x_1$ |
| $z_0 + z_1 = \mathrm{texpm1p}(x_0 + x_1)$ | Twofold $e^x - 1$ of coupled argument $x_0 + x_1$ |
| $z_0 + z_1 = \mathrm{pexpm1}(x_0 + x_1)$ | Coupled $e^x - 1$ of coupled argument $x_0 + x_1$ |

For plain C, add suffix "f" to function name if `float` type. C++ interface would support type overloading.

Term "dotted" means ordinary floating-point number, in contract to "shaped" like twofold or coupled.

Recalling from [1-2], term "coupled" means special case of renormalized twofold, such that rounding $x_0 + x_1$ to 1x-precision gives exactly $x_0$. Particularly this means that $x_1$ is very small comparing $x_0$ by magnitude. Coupled are similar to double-length numbers by Dekker [7] and to double-doubles [5-6].

Here I do not define any special algorithm for twofold functions of coupled argument.

Function $z_0 + z_1 = \text{texp0}(x_0)$ should assess accuracy of $\exp(x_0)$ from C/C++ standard library, so must guarantee $z_0 = \exp(x_0)$ bitwise. This may require additional care and negatively impact performance. Use faster functions $\text{pexp0}(x_0)$ and $\text{pexpm10}(x_0)$ if you do not need bitwise reproducibility.

These prefixed with "p" functions include fast renormalizing result to ensue $z_0 + z_1$ is "coupled". This step takes only 3 dotted add/subtract operations, so is quite fast. See renormalization details in [1].

Following is explicit pexp0/pexpm10 algorithms:

<u>Algorithm</u>: $z_0 + z_1 = \text{pexp0}(x_0)$
(1) If $x_0 < \ln 2^{\min}$, then let $z_0 = z_1 = 0$
(2) If $x_0 > \ln 2^{\max}$, then let $z_0 = z_1 = +\infty$
(3) Otherwise, if $\ln 2^{\min} \le x_0 \le \ln 2^{\max}$, then:
   (a) Decompose exactly $x_0 = 2^L m + 2^{-K} n + y$
   (b) Get $E \approx e^{2^L m}$ and $C \approx e^{2^{-K} n}/N!$ from table
   (c) Compute $T \approx e^y N!$ With Horner scheme
   (d) Compute $z_0 + z_1$ like $E \cdot (C \cdot T)$
(4) Renormalize fast $z_0 + z_1$

<u>Algorithm</u>: $z_0 + z_1 = \text{pexpm10}(x_0)$
(1) If $|x_0| > \ln 2$, let $z_0 + z_1 = \text{pexp0}(x_0) - 1$
(2) Otherwise, if $|x_0| \le \ln 2$, then:
   (a) Decompose exactly $x_0 = 2^{-K} n + y$
   (b) Let $c = n/2^K$, and get $C \approx e^c - 1$ from table
   (c) Compute $T \approx e^y - 1$ With Horner scheme
   (d) Compute $z_0 + z_1$ like $CT + (C + T)$
(3) Renormalize fast $z_0 + z_1$

Algorithm parameters depending on floating-point format:

| | $\text{texp0}(x)$ | $\text{texpm10}(x)$ |
|---|---|---|
| binary64 (`double`) | $L = 3, K = 5, N = 12, \min = -1074, \max = 1024$ | $K = 7, N = 10$ |
| binary32 (`float`) | $L = 1, K = 5, N = 6, \min = -149, \max = 128$ | $K = 7, N = 5$ |

Ensure bitwise reproducing standard functions:

<u>Algorithm</u>: $z_0 + z_1 = \text{texp0}(x_0)$
(1) Twofold $v_0 + v_1 = \text{pexp0}(x_0)$ -- may omit renormalizing $v_0 + v_1$
(2) Dotted $z_0 = \exp(x_0)$
(3) Dotted $z_1 = (v_0 - z_0) + v_1$

<u>Algorithm</u>: $z_0 + z_1 = \text{texpm10}(x_0)$
(1) Twofold $w_0 + w_1 = \text{pexpm10}(x_0)$ -- may omit renormalizing $w_0 + w_1$
(2) Dotted $z_0 = \text{expm1}(x_0)$
(3) Dotted $z_1 = (w_0 - z_0) + w_1$

Twofold functions of twofold argument:

<u>Algorithm</u>: $z_0 + z_1 = \text{texp}(x_0 + x_1)$
(1) Twofold $v_0 + v_1 = \text{pexp0}(x_0)$ -- may omit renormalizing $v_0 + v_1$

(2) Dotted $z_0 = \exp(x_0)$
(3) Dotted $t_1 = \mathrm{expm1}(x_1)$
(4) Dotted $z_1 = z_0 t_1 + \big(v_1 + (v_0 - z_0)\big)$

Algorithm: $z_0 + z_1 = \mathrm{texpm1}(x_0 + x_1)$
(1) Twofold $w_0 + w_1 = \mathrm{pexpm10}(x_0)$ -- may omit renormalizing
(2) Dotted $z_0 = \mathrm{expm1}(x_0)$
(3) Dotted $t_1 = \mathrm{expm1}(x_1)$
(4) Let $z_1 = (z_0 + 1)t_1 + \big(w_1 + (w_0 - z_0)\big)$

Subtracting $v_0 - z_0$ may require special care to avoid $z_1$ getting NaN if $z_0 = v_0$ is infinity (same for $w_0$).

Special functions $\mathrm{texpp}(x_0 + x_1)$ and $\mathrm{texpm1p}(x_0 + x_1)$ for coupled arguments do not use specifics (coupled-ness) of argument and just call $\mathrm{texp}(x_0 + x_1)$ and $\mathrm{texpm1}(x_0 + x_1)$.

Functions accepting and returning special "coupled" kind of twofold. Fast renormalization of result is appropriate here because $|z_0| \geq |z_1|$. See renormalization details in [1]:

Algorithm: $z_0 + z_1 = \mathrm{pexp}(x_0 + x_1)$
(1) Let $z_0 + z_1 = \mathrm{texpp}(x_0 + x_1)$
(2) Renormalize fast $z_0 + z_1$

Algorithm: $z_0 + z_1 = \mathrm{pexpm1}(x_0 + x_1)$
(1) Let $z_0 + z_1 = \mathrm{texpm1p}(x_0 + x_1)$
(2) Renormalize fast $z_0 + z_1$

## Vectoring for SIMD

Basic algorithms for $\mathrm{pexp0}(x_0)$ and $\mathrm{pexpm10}(x_0)$ unfortunately include if-then-else branching by value of $x_0$, which branching is not good for single-instruction-multiple-data (SIMD) programing.

However, we still could vector these algorithms conditionally:

- Given vectored argument $x_{01}, \ldots, x_{0I}$, execute vectored if all $x_i$ fit function's main interval, that is $\ln 2^{\min} < x_{0i} < \ln 2^{\max}$ for $\mathrm{texp0}(x_0)$, and $\ln 2 < x_{0i} < \ln 2$ for $\mathrm{pexpm10}(x_0)$
- Otherwise, simulate vectored interface by evaluating for each $x_{0i}$ sequentially

For $\mathrm{pexp0}(x_0)$ this usually would call parallelized (vectored) variant assuming user code is reasonable and avoids exponentiation of too large arguments. For $\mathrm{texpm10}(x_0)$, this also must call parallel code in most cases, assuming programmers use $\mathrm{expm1}(x_0)$ adequately, only if $x_0$ is presumably small.

Future SIMD processors, like AVX-512 announced by Intel, would support masked SIMD operations, like `_mm512_mask_`***`operation`***`_pd()` for example. This must allow more flexible vectoring of twofolds.

Other twofold algorithms like $\mathrm{texp}(x_0 + x_1)$ and $\mathrm{texpm1}(x_0 + x_1)$ do not include branching. We could directly vector them for SIMD, if vectored $\exp(x_0)$ and $\mathrm{expm1}(x_0)$ were available.

# Twofold logarithm

- Newton inversion
- Algorithm summary
- Vectoring for SIMD

## Newton inversion

Let us compute twofold logarithm by Newton inversion of twofold exponent. Standard 1x-precision logarithm from C/C++ math library can provide very good initial approximation, so iterations would converge very quickly. Actually, just one iteration is enough.

Given equation $y = \mathrm{f}(x)$ and initial guess $y \approx \mathrm{f}(x_0)$, let us assess $x_1 = x - x_0 \approx (\mathrm{f}(x) - \mathrm{f}(x_0))/\mathrm{f}'(x_0)$. This is form of Taylor series $\mathrm{f}(x) = \mathrm{f}(x_0) + \mathrm{f}'(x_0)\,(x - x_0) + R$, where small $R = (x - x_0)^2\,\mathrm{f}''(\xi)/2$ for some $\xi$ found between $x$ and $x_0$.

Let $a = \mathrm{f}^{-1}(y)$ be exact solution, and $\Delta x_0 = a - x_0$ be deviation of $x_0$, and $\Delta x = a - x$ deviation of $x$. Then $\Delta x = -(\Delta x_0)^2\,\mathrm{f}''(\xi)/2\,\mathrm{f}'(\xi)$. So twofold $x = x_0 + x_1$ is 2x-precise, if initial guess $x_0$ is 1x-precise and $\mathrm{C}(\xi) = \mathrm{f}''(\xi)/2\,\mathrm{f}'(\xi)$ does not exceed 1 by magnitude.

If function $\mathrm{f}(x)$ is $e^x$ or $e^x - 1$, then $\mathrm{C}(\xi)$ always equals 1/2. Thus one Newton iteration is enough for twofold logarithm, assuming 1x-precision logarithm functions from C/C++ standard math library supply accurate initial guess for $\ln(y)$ and $\ln(1+y)$.

* * *

Iteration formulas for inverting $\mathrm{f}(x) = e^x$ is following. Given twofold $y = y_0 + y_1$ and initial guess $x_0$, compute $x_1 = (\mathrm{f}(x) - \mathrm{f}(x_0))/\mathrm{f}'(x_0)$ like $(y - e^{x_0})/e^{x_0} = ye^{-x_0} - 1$. Because $ye^{-x_0}$ is very close to 1, we need to compute $ye^{-x_0}$ with at least 2x-precision like twofold $(y_0 + y_1) \cdot \mathrm{pexp0}(-x_0)$.

Three problems with this direct formula:

- Initial guess like $x_0 = \log(y_0)$ may get not enough accurate if $y_0 + y_1$ is close to 1
- Twofold $t_0 + t_1 = \mathrm{pexp0}(-x_0)$ may get inaccurate if $y_0 \approx 1$ and therefore $t_0 \approx 1$
- Formula may get inaccurate if $y_0 + y_1$ is too large or, conversely, is too close to 0

Let us use following tricks to address these problems:

- If $y_0 < 1/2$ or $y_0 > 2$, decompose $y = 2^n z$ and compute $x_0 + x_1$ like $n \ln 2 + \ln z$
- If $1/2 \le y_0 \le 2$, then let initial guess be $x_0 = \mathrm{log1p}\big((y_0 - 1) + y_1\big)$, and compute $x_1$ from $(y_0 + y_1) \cdot (s_0 + s_1 + 1) - 1$ reducing this to $ys + (y - 1)$ where $s_0 + s_1 = \mathrm{pexpm10}(-x_0)$

Note that $y_0 - 1$ is exact by Sterbenz lemma, so $(y_0 - 1) + y_1$ holds the most significant bits of $y - 1$.

Note that this algorithm cannot guarantee if $x_0 = \log(y_0)$, so bitwise reproducing of standard function would require additional corrective steps.

* * *

Inverting $\mathrm{f}(x) = e^x - 1$ is similar. Newton formula for $x_1$ is $(y - e^{x_0} + 1)/e^{x_0} = (1 + y)e^{-x_0} - 1$. Again, $(1 + y)e^{-x_0}$ is close to 1, so we need at least 2x-precision like $\big(1 + (y_0 + y_1)\big) \cdot \mathrm{pexp0}(-x_0)$.

Similar problems with this direct formula:

- Twofold $t_0 + t_1 = \mathrm{pexp0}(-x_0)$ may get inaccurate if $y_0 \approx 0$ and therefore $t_0 \approx 1$
- Formula may get inaccurate if $y_0 + y_1$ is too large or, conversely, is too close to $-1$

Following tricks would address these problems:

- If $y_0 < -1/2$ or $y_0 > 1$, compute $x_0 + x_1$ as twofold $\log(1 + y)$ like described above

- If $-1/2 \le y_0 \le 1$, let initial guess be $x_0 = \text{log1p}(y_0)$ and $s_0 + s_1 = \text{pexpm10}(-x_0)$, and compute $x_1$ from $(y_0 + y_1 + 1) \cdot (s_0 + s_1 + 1) - 1$ reducing this formula to $ys + y + s$

Again, this algorithm cannot guarantee if $x_0 = \text{log1p}(y_0)$ if $y$ is large or close to $-1$, so reproducing bitwise would require additional care.

* * *

These algorithms work fine only in special case if twofold $y_0 + y_1$ is "coupled", that is if $y_1$ is very minor comparing $y_0$, and 1x-precision standard library can provide good initial guess for Newton iterations.

In general case, let us renormalize twofold $y_0 + y_1$ into coupled $v_0 + v_1$ with exactly equal value, and compute twofold logarithm $u_0 + u_1$ with one of these algorithms above. Then deduce $x_0 + x_1$ like:

- Compute $x_0$ like $\log(y_0)$ with 1x-precision library , or $\text{log1p}(y_0)$ accordingly
- Compute $x_1$ with 1x-precision expression $(u_0 - x_0) + u_1$

Parentheses in $(u_0 - x_0) + u_1$ are important. If occasionally $u_0 \approx x_0$, subtracting $u_0 - x_0$ is exact by Sterbenz lemma (see [8]), otherwise $u_1$ is almost negligible.

## Algorithm summary

Here let me enlist all functions I propose for twofold logarithm, and explicitly write algorithms for them.

| Function | Description |
|---|---|
| $x_0 + x_1 = \text{plog0}(y_0)$ | Twofold logarithm of dotted argument, $y_1 = 0$ |
| $x_0 + x_1 = \text{tlog0}(y_0)$ | Twofold logarithm of dotted $y_0$, ensure $x_0 = \log(y_0)$ |
| $x_0 + x_1 = \text{tlogp}(y_0 + y_1)$ | Twofold logarithm of coupled argument, $\lvert y_1\rvert \ll \lvert y_0\rvert$ |
| $x_0 + x_1 = \text{tlog}(y_0 + y_1)$ | Twofold logarithm of twofold argument, any $y_1$ |
| $x_0 + x_1 = \text{plog}(y_0 + y_1)$ | Coupled logarithm of coupled argument |
| $x_0 + x_1 = \text{plog1p0}(y_0)$ | Twofold $\ln(1 + y)$ of dotted argument, $y_1 = 0$ |
| $x_0 + x_1 = \text{tlog1p0}(y_0)$ | Twofold $\ln(1 + y)$ of dotted $y_0$, ensure $x_0 = \text{log1p}(y_0)$ |
| $x_0 + x_1 = \text{tlog1pp}(y_0 + y_1)$ | Twofold $\ln(1 + y)$ of coupled argument, $\lvert y_1\rvert \ll \lvert y_0\rvert$ |
| $x_0 + x_1 = \text{tlog1p}(y_0 + y_1)$ | Twofold $\ln(1 + y)$ of twofold argument, any $y_1$ |
| $x_0 + x_1 = \text{plog1p}(y_0 + y_1)$ | Coupled $\ln(1 + y)$ of coupled argument |

To recall, term "dotted" means ordinary floating-point number, not twofold. Term "coupled" means special case of twofold, such that rounding of $y = y_0 + y_1$ to 1x-precision gives exactly $y_0$.

Simplest case if $y_1 = 0$. Note that we execute some steps with twofold/coupled precision:

<u>Algorithm</u>: $x_0 + x_1 = \text{plog0}(y_0)$

(1) Decompose $y_0 = 2^n z_0$
  (a) If $y_0 < 1/2$ or $y_0 > 2$, let $z_0 = \text{frexp}(y_0, \&n)$
  (b) Otherwise, let $z_0 = y_0$ and $n = 0$
(2) Compute $r_0 + r_1$ for $z_0$
  (a) Dotted $r_0 = \log(z_0)$
  (b) Twofold $s_0 + s_1 = \text{pexpm10}(-r_0)$ -- may omit renormalizing $s_0 + s_1$
  (c) Twofold $u_0 + u_1 = zs + (z - 1)$
  (d) Dotted $r_1 = u_0 + u_1$
(3) Twofold $x_0 + x_1 = (r_0 + r_1) + n \ln 2$
(4) Renormalize fast $x_0 + x_1$

<u>Algorithm</u>: $x_0 + x_1 = \text{plog1p0}(y_0)$

(1) If $y_0 < -1/2$ or $y_0 > 1$, then:
  (a) Let $v_0 + v_1$ be renormalized $1 + y_0$

(b) Let $x_0 + x_1 = \text{plog}(v_0 + v_1)$
(2) If $-1/2 \le y_0 \le 1$, then:
   (a) Dotted $x_0 = \text{log1p}(y_0)$
   (b) Twofold $s_0 + s_1 = \text{pexpm10}(-r_0)$ -- may omit renormalizing $s_0 + s_1$
   (c) Twofold $u_0 + u_1 = (1 + y)s + y$
   (d) Dotted $x_1 = u_0 + u_1$
(3) Renormalize fast $x_0 + x_1$

Note that by Sterbenz lemma, $v_1 = 0$ in step (2.a) of last algorithm if $y_0$ between -1 and -1/2.

Ensure bitwise reproducing of standard C/C++ library functions:

Algorithm: $x_0 + x_1 = \text{tlog0}(y_0)$
(1) Compute $x_0 + x_1 = \text{plog0}(y_0)$ -- may omit renormalizing $x_0 + x_1$
(2) Make correction to ensure $x_0 = \log(y_0)$ bitwise:
   (a) Bitwise $u_0 = \log(y_0)$
   (b) Replace $x_1 = x_1 + (u_0 - x_0)$
   (c) Replace $x_0 = u_0$

Algorithm: $x_0 + x_1 = \text{tlog1p0}(y_0)$
(3) Compute $x_0 + x_1 = \text{plog1p0}(y_0)$ -- may omit renormalizing $x_0 + x_1$
(4) Make correction to ensure $x_0 = \log(y_0)$ bitwise:
   (a) Bitwise $u_0 = \text{log1p}(y_0)$
   (b) Replace $x_1 = x_1 + (u_0 - x_0)$
   (c) Replace $x_0 = u_0$

Coupled logarithm of coupled argument:

Algorithm: $x_0 + x_1 = \text{plog}(y_0 + y_1)$
(1) Decompose $y_0 + y_1 = 2^n(z_0 + z_1)$
   (a) If $y_0 < 1/2$ or $y_0 > 2$, then:
      - Let $z_0 = \text{frexp}(y_0, \&n)$
      - Let $z_1 = \text{ldexp}(y_1, -n)$
   (c) Otherwise, let $z_0 + z_1 = y_0 + y_1$ and $n = 0$
(2) Compute $r_0 + r_1$ for $z_0 + z_1$
   (a) Dotted $r_0 = \text{log1p}\big((z_0 - 1) + z_1\big)$
   (b) Twofold $s_0 + s_1 = \text{pexpm10}(-r_0)$ -- may omit renormalizing $s_0 + s_1$
   (c) Twofold $u_0 + u_1 = zs + (z - 1)$
   (d) Dotted $r_1 = u_0 + u_1$
(3) Twofold $x_0 + x_1 = (r_0 + r_1) + n \ln 2$
(4) Renormalize fast $x_0 + x_1$

Algorithm: $x_0 + x_1 = \text{plog1p}(y_0 + y_1)$
(1) If $y_0 < -1/2$ or $y_0 > 1$, then:
   (a) Let $x_0 + x_1 = \text{tlogp}(1 + y)$
(2) If $-1/2 \le y_0 \le 1$, then:
   (a) Dotted $x_0 = \text{log1p}(y_0)$
   (b) Twofold $s_0 + s_1 = \text{pexpm10}(-r_0)$ -- may omit renormalizing $s_0 + s_1$
   (c) Twofold $u_0 + u_1 = ys + y + s$
   (d) Dotted $x_1 = u_0 + u_1$
(3) Renormalize fast $x_0 + x_1$

Twofold functions of coupled argument ensure $x_0$ bitwise reproduces standard library's result:

Algorithm: $x_0 + x_1 = \mathrm{tlogp}(y_0 + y_1)$

(1) Twofold $x_0 + x_1 = \mathrm{plog}(y_0 + y_1)$ -- may omit renormalizing $x_0 + x_1$
(2) Make correction to ensure $x_0 = \log(y_0)$ bitwise:
    (a) Compute $u_0 = \log(y_0)$
    (b) Replace $x_1 = x_1 + (u_0 - x_0)$
    (c) Replace $x_0 = u_0$

Algorithm: $x_0 + x_1 = \mathrm{tlog1pp}(y_0 + y_1)$

(1) Compute $x_0 + x_1 = \mathrm{plog1p}(y_0)$ -- may omit renormalizing $x_0 + x_1$
(2) Make correction to ensure $x_0 = \log(y_0)$ bitwise:
    (d) Bitwise $u_0 = \mathrm{log1p}(y_0)$
    (e) Replace $x_1 = x_1 + (u_0 - x_0)$
    (f) Replace $x_0 = u_0$

Twofold logarithm of arbitrary (not necessarily renormalized) twofold argument:

Algorithm: $x_0 + x_1 = \mathrm{tlog}(y_0 + y_1)$

(1) Twofold $v_0 + v_1$ be renormalized $y_0 + y_1$
(2) Twofold $u_0 + u_1 = \mathrm{plog}(v_0 + v_1)$ -- may omit renormalizing $u_0 + u_1$
(3) Dotted $x_0 = \log(y_0)$
(4) Dotted $x_1 = u_1 + (u_0 - x_0)$

Algorithm: $x_0 + x_1 = \mathrm{tlog1p}(y_0 + y_1)$

(1) Twofold $v_0 + v_1$ be renormalized $y_0 + y_1$
(2) Twofold $u_0 + u_1 = \mathrm{plog1p}(v_0 + v_1)$ -- may omit renormalizing $u_0 + u_1$
(3) Dotted $x_0 = \mathrm{log1p}(y_0)$
(4) Dotted $x_1 = u_1 + (u_0 - x_0)$

Note that these algorithms do not need to check argument $y_0 + y_1$ for domain error:

- If $y_0$ does not fit domain, then $x_0$ is NaN due to calling $\log(y_0)$ or $\mathrm{log1p}(y_0)$
- For coupled, $y_0 + y_1$ missing domain implies $y_0$ misses, so again $x_0$ is NaN
- For general case, if $y_0 + y_1$ does not fit, then $v_0$ does not fit, so $x_1$ is NaN

General-case result may look tricked, if $x_0$ is a number but $x_1$ is NaN. This is correct behavior, as we need main part of twofold result to reproduce standard 1x-precision result on the bitwise basis.

## Vectoring for SIMD

Logarithm basic algorithms for dotted argument $y_0$ or coupled $y_0 + y_1$, include if-then-else branching if value of $y_0$ fits interval [ ½,2] or [-½,1], which branching is not good for SIMD computations.

However, compromise is still possible, if we combine true SIMD with simulating SIMD via several calls of sequential subroutines. Given vectored input $y_{01}, \dots, y_{0\mathrm{I}}$, we might evaluate with true SIMD if branching decision is same for every $y_{0i}$, or simulate SIMD otherwise. There is a chance that true SIMD would work often, so average performance would be significantly better than for purely sequential computations.

Some processors like announced Intel Skylake would support conditional (masked) SIMD operations like for example `_mm512_mask_`***`operation`***`_pd()`, which must allow more flexible approaches.

Other twofold logarithm algorithms do not include if-then-else branching so allow vectoring for SIMD.

Of course, for vectoring we need simulated or truly SIMD variant of `tlogp()` and `tlog1pp()` and of `texp0()`. Vectoring would also require SIMD variant of standard functions `log()` and `log1p()`.

# Experimental code

- C/C++ interface
- Implementation
- Demo examples
- SIMD extension

## C/C++ interface

With this article, I provide experimental code implementing twofold exponent and logarithm functions. Here I describe plain C and C++ interfaces for these functions, aligned with twofold arithmetic interface.

In paper [2] entitled "Twofolds for C and C++", I propose plain C interface for maximal use of processor registers for parameters of twofold operations. For example:

```
#include "twofold.h"
float x0,x1, y0,y1, z0,z1;   // twofold x0+x1, y0+y1, and z0+z1
z0 = taddf(x0,x1,y0,y1,&z1); // z0+z1 is sum of x0+x1 and y0+y1
```

Here, prefix "t" in function name means twofold, "add" means summation, and suffix "f" means `float`. Compiler transfers returned value and majority of parameters via CPU registers for x86 processors if in 64-bits mode, so ensuring maximal performance.

Using registers for calling slower functions is less beneficial. However, twofold functions follow same scheme in order to unify interfaces look-and-feel. For example:

```
#include "texplog.h"
float x0,x1, y0,y1, z0,z1; // twofold x0+x1, y0+y1, z0+z1
z0 = texpf(x0,x1, &z1);    // z0+z1 is exponent  of x0+x1
x0 = tlogf(y0,y1, &x1);    // x0+x1 is logarithm of y0+y1
```

C++ interface additionally allows type polymorphism, so you can omit suffix "f":

```
#include "texplog.h"
float x0,x1, y0,y1, z0,z1;
z0 = texp(x0,x1, &z1); // note: no suffix "f" in function name
x0 = tlog(y0,y1, &x1);
```

On top of that, C++ interface defines `twofold<T>` generic types where T is `float` or `double`, and functions. These convenience types and functions belong to "tfcp" namespace. For example:

```
#include "twofold.h"
#include "texplog.h"
using namespace tfcp;
twofold<float> x, y, z;
z = texp(x); // z is twofold exponent  of x
x = tlog(y); // x is twofold logarithm of y
```

Finally, C++ convenience interface overloads standard functions exp(x) ad log(y). For twofold x and y, these functions would imply twofold operations texp(x) and tlog(y). For example:

```
#include "twofold.h"
#include "texplog.h"
using namespace tfcp;
twofold<float> x, y, z;
z = exp(x); // same as z=texp(x)
x = log(y); // same as x=tlog(y)
```

Following is summary of plain C interface for twofold/coupled exponent and logarithm:

| | Type `float` | Type `double` |
|---|---|---|
| Exp | `z0=pexp0f(x0    ,&z1)`<br>`z0=texp0f(x0    ,&z1)`<br>`z0=texppf(x0,x1,&z1)`<br>`z0=pexpf (x0,x1,&z1)`<br>`z0=pexpf (x0,x1,&z1)` | `z0=pexp0(x0    ,&z1)`<br>`z0=texp0(x0    ,&z1)`<br>`z0=texpp(x0,x1,&z1)`<br>`z0=pexp (x0,x1,&z1)`<br>`z0=pexp (x0,x1,&z1)` |
| Expm1 | `z0=pexpm10f(x0    ,&z1)`<br>`z0=texpm10f(x0    ,&z1)`<br>`z0=texpm1pf(x0,x1,&z1)`<br>`z0=pexpm1f (x0,x1,&z1)`<br>`z0=pexpm1f (x0,x1,&z1)` | `z0=pexpm10(x0    ,&z1)`<br>`z0=texpm10(x0    ,&z1)`<br>`z0=texpm1p(x0,x1,&z1)`<br>`z0=pexpm1 (x0,x1,&z1)`<br>`z0=pexpm1 (x0,x1,&z1)` |
| Log | `x0=plog0f(y0    ,&x1)`<br>`x0=tlog0f(y0    ,&x1)`<br>`x0=tlogpf(y0,y1,&x1)`<br>`x0=plogf (y0,y1,&x1)`<br>`x0=plogf (y0,y1,&x1)` | `x0=plog0(y0    ,&x1)`<br>`x0=tlog0(y0    ,&x1)`<br>`x0=tlogp(y0,y1,&x1)`<br>`x0=plog (y0,y1,&x1)`<br>`x0=plog (y0,y1,&x1)` |
| Log1p | `x0=plog1p0f(y0    ,&x1)`<br>`x0=tlog1p0f(y0    ,&x1)`<br>`x0=tlog1ppf(y0,y1,&x1)`<br>`x0=plog1pf (y0,y1,&x1)`<br>`x0=plog1pf (y0,y1,&x1)` | `x0=plog1p0(y0    ,&x1)`<br>`x0=tlog1p0(y0    ,&x1)`<br>`x0=tlog1pp(y0,y1,&x1)`<br>`x0=plog1p (y0,y1,&x1)`<br>`x0=plog1p (y0,y1,&x1)` |

Programming in C++ you may omit suffix “f”, so function names are uniform for `float` and `double`.

Summary of C++ convenience interface. Recall that twofold<T> is structure of two fields named as “value” and “error”. Function names are uniform for type T be `float` or `double`:

| | **Detailed** | **Standard-like** |
|---|---|---|
| Exp | `z = pexp0(x.value)`<br>`z = texp0(x.value)`<br>`z = texpp(x)`<br>`z = texp (x)`<br>`z = pexp (x)` | `z = exp(x)` |
| Expm1 | `z = pexpm10(x.value)`<br>`z = texpm10(x.value)`<br>`z = texpm1p(x)`<br>`z = texpm1 (x)`<br>`z = pexpm1 (x)` | `z = expm1(x)` |
| Log | `x = plog0(y.value)`<br>`x = tlog0(y.value)`<br>`x = tlogp(y)`<br>`x = tlog (y)`<br>`x = plog (y)` | `x = log(y)` |
| Log1p | `x = plog1p0(y.value)`<br>`x = tlog1p0(y.value)`<br>`x = tlog1pp(y)`<br>`x = tlog1p (y)`<br>`x = plog1p (y)` | `x = log1p(y)` |

Standard-like exp(x) and log(y) calling twofold (not coupled) ensures “value” parts of x and z reproduce C/C++ library functions bitwise, so “error” parts can assess inaccuracy of standard-math code.

## Implementation

If you would like to explore my experimental implementation, this subsection explains code structure.

Code is available as code.zip archive downloadable from project Web site, see link in How to download section. Code is free for non-commercial and academic use, presumably for learning/investigating. Note that code quality may be not good enough for any commercial or mission-critical applications.

Zip archive contains all necessary sources and make files, and does not depend on any other packages. All you need for trying it is C/C++ compiler(s). I designed and tested code for Microsoft and GNU C/C++ compilers, versions Visual Studio 2013 Express, and Cygwin 4.8.3, both for Windows on x86-compatible processor. Theoretically, code may work in 32-bits mode, but I tested only x86-64 mode.

Note that performance of twofold arithmetic critically depends on fast fused-multiply-add (FMA). Thus, you would need processor that supports FMA, like Intel Haswell, or newer like Broadwell. Theoretically, code may work with compatible AMD processors, but I tested only Intel Haswell.

To explore code, unpack code.zip into any directory at your computer. Twofold exponent and logarithm implementation locates under `code/texplog` folder. Following is folders full list:

```
code/
    auxiliary
    corner_cases
    examples
    lups
    perfmath
    perftest
    texplog
    twofold
    xblas
```

Twofold arithmetic implemented as "`twofold.h`" found under `code/twofold` folder. Exponent and logarithm defined as "`texplog.h`" and implemented with "`texplog.c`" found under `code/texplog`. Other folders contain tests and auxiliary files, specifically:

- auxiliary: folder w/auxiliary files, random numbers generator, and timer for performance tests
- corner_cases, examples, lups, xblas: examples of using twofold arithmetic, in linear algebra etc.
- permath, perftest: performance tests for C/C++ standard math library, and twofold arithmetic

Root folder contains master make file that runs all tests, supports comprehensive testing (~30-40 min) or fast testing (~3-5 min). Fast testing must work on any processor supporting SSE2. Comprehensive script additionally tests code specific for AVX and for AVX+FMA. Try with ordinary make or Microsoft nmake, like following:

```
cd …/code
make gcc_fast
make gcc
nmake cl
```

Each subfolder holds similar master make file, plus maybe one or more partial make files, for example:

```
cd texplog
make gcc –f texplog_demo.mk
```

Folder `code/texplog` contains following partial make files:

- taylor.mk -- generates auxiliary tables, shows accuracy of Taylor series
- texplog_const.mk -- generates tables of constants used by texp() and tlog()
- texplog_demo.mk -- simple demo, BTW tests corner cases, ±inf and nan

- texplog_link.mk -- link test for texplog.h
- texplog_perf.mk -- performance test
- texplog_test.mk -- accuracy test

Please consider these make files as samples, modify if necessary to adjust for your test environment.

## Demo examples

In particular, demo examples illustrate bitwise reproducibility of standard functions:

```
// Call right type of exp() et al:
exp1   = std_exp  ((T) 1);
expm11 = std_expm1((T) 1);
log2   = std_log  ((T) 2);
log1p1 = std_log1p((T) 1);

// Bitwise reproducibility:
CHECK((r = exp  (  unity)).value == exp1  );
CHECK((r = expm1(  unity)).value == expm11);
CHECK((r = log  (2*unity)).value == log2  );
CHECK((r = log1p(  unity)).value == log1p1);
```

Note that `double` and `float` types of same standard function like exp() may return different results.

## SIMD extension

SIMD extension for twofold exponent and logarithm functions is not available yet.

However, twofolds code depends on hardware fast-FMA available only with newer processors w/SIMD. See Older CPUs and Compiler flags sections, which explains compiling for older and newer processors.

# Analysis

- Performance
- Accuracy

## Performance

Performance test uses `texplog_perf.mk` make file found under `code/texplog` folder, where you may also find detailed test logs and Excel table summarizing testing results. Here I describe test environment and analyze results.

For testing, I used my laptop: HP Pavilion 15, with Intel Core i5-4200U (Haswell) processor. Processor frequency is flexible and depends on actual workload. In my testing, frequency was around 2.55 GHz, according to my observation with Windows Task Manager.

I tested with two compilers, Microsoft and GNU; here I describe GNU results. I used GNU gcc/g++ 4.8.3 from Cygwin package, 64-bit variant, which you can download from Cygwin web site:

http://cygwin.org/

With this compiler version, GNU standard math library performance looks like following, according to my test found under `code/perfmath` folder. Performance is measured in millions of function calls per second. Column "quad" means `__float128` type with `libquadmath` as supported by GNU compiler.

Results for scalar code, without vectoring for SIMD:

Performance (MOPS):

| | float | double | quad |
|---|---|---|---|
| exp | 50.0152 | 48.6687 | 0.768865 |
| expm1 | 68.923 | 64.4572 | 0.607182 |
| log | 51.2633 | 41.6789 | 0.763506 |
| log1p | 65.7593 | 59.1496 | 0.50567 |

At frequency around 2.5 GHz, performance of `float` and `double` functions near 50 MOPS implies about 50 processor clock "ticks" per function call. Quad-precision results appear 50-100 times worse than that. Twofold functions should take the niche in between, be at least ~10x times faster than quad-precision.

Following results for twofold exponent show that my experimental implementation does fit this goal:

Performance (MOPS):

| | float | double | | float | double |
|---|---|---|---|---|---|
| pexp0 | 24.7945 | 18.1997 | pexpm10 | 22.9724 | 17.1882 |
| texp0 | 13.8971 | 11.8716 | texpm10 | 16.6896 | 14.4446 |
| texpp | 13.6488 | 11.8232 | texpm1p | 13.9798 | 12.2363 |
| texp | 13.7022 | 11.8582 | texpm1 | 13.9394 | 12.2799 |
| pexp | 12.6837 | 11.1004 | pexpm1 | 13.2497 | 11.4258 |

As expected, basic coupled-precision function $t_0 + t_1 = \mathrm{pexp}(x_0)$ of `double` argument $x_0$ takes ~120 processor ticks per function call, so operates at nearly 20 million calls pes second on 2.5 GHz processor. Function $t_0 + t_1 = \mathrm{texp}(x_0 + x_1)$ operates at ~12 MOPS, so outperforms quad-precision by 15+ times. Function $s_0 + s_1 = \mathrm{texpm1}(x_0 + x_1)$ operates at 12+ MOPS and outperforms quad by 20+ times.

Following are my testing results for twofold logarithm:

Performance (MOPS):

| | float | double | | float | double |
|---|---|---|---|---|---|
| plog0 | 12.2482 | 9.9473 | plog1p0 | 7.5686 | 6.61857 |
| tlog0 | 10.579 | 8.96292 | tlog1p0 | 6.92869 | 6.02879 |
| tlogp | 8.38499 | 7.16246 | tlog1pp | 8.56824 | 7.15633 |
| tlog | 8.18446 | 7.00466 | tlog1p | 8.40729 | 7.06084 |
| plog | 10.2372 | 8.41465 | plog1p | 9.83094 | 8.34681 |

Twofold function $\mathrm{tlog}(y_0 + y_1)$ is not as fast with its 7 MOPS, but it still outperforms quad-precision $\mathrm{logq}(y)$ by 9.2 times, so nearly meets the 10x performance goal. Function $\mathrm{tlog1p}(y_0 + y_1)$ shows 7.1 MOPS and so outperforms $\mathrm{log1pq}(y)$ by 14+ times, significantly exceeding the 10x goal.

Outperforming quad-precision is not free; this is balance with some loss of accuracy. See details below.

Twofold functions performance over `float` type is much worse than standard functions of `double`, as expected. Obviously, it is better using standard `double` for measuring inaccuracy of `float` calculations.

## Accuracy

Accuracy test locates under the same `code/texplog` folder and uses the `texplog_test.mk` make file.

The test evaluates each of the twofold exponent and logarithm functions and compares with higher-precision etalon function. C/C++ standard functions of `double` type are etalon for twofold over `float`, and GNU quad-math function of `__float128` type are etalon for twofolds over `double`.

The test tries each function against ~1 million of random samples simulating 2x-precise arithmetic. The test checks maximal and average deviation from etalon. Sort of $L_0$ and $L_1$ norms, though not quite: the test allows a few (≤2 per million samples) “warnings” if deviation exceeds $L_0$ threshold in corner cases.

My reasoning for such looser criteria is important, let me emphasize it in Conclusion subsection below.

The acceptance thresholds for the $L_0$ and $L_1$ criteria were the following, in terms of relative error:

- Average ($L_1$) for `double` type: $2^{-100}$ ($\approx 7.9 \cdot 10^{-31}$)
- Average ($L_1$) for `float` type: $2^{-42}$ ($\approx 2.3 \cdot 10^{-13}$)
- Maximal ($L_0$) for `double` type: $2^{-95}$ ($\approx 2.5 \cdot 10^{-29}$)
- Maximal ($L_0$) for `float` type: $2^{-38}$ ($\approx 3.6 \cdot 10^{-12}$)

Unfortunate exception is twofold logarithm, function $\mathrm{tlog}(y_0 + y_1)$ et al, which are less accurate in $L_0$ norm by approximately 2 bits. Thresholds for logarithm were:

- Average ($L_1$) for `double` type: $2^{-98}$ ($\approx 3.2 \cdot 10^{-30}$)
- Average ($L_1$) for `float` type: $2^{-42}$ ($\approx 2.3 \cdot 10^{-13}$)
- Maximal ($L_0$) for `double` type: $2^{-93}$ ($\approx 1.0 \cdot 10^{-28}$)
- Maximal ($L_0$) for `float` type: $2^{-36}$ ($\approx 1.5 \cdot 10^{-11}$)

Over `double` type, given a twofold result like $z_0 + z_1 = \mathrm{texp}(x_0 + x_1)$, average accuracy of $z_0 + z_1$ like 100+ bits means that $z_1$ assesses deviation $\Delta z_0 = e^{x_0} - z_0$ of standard result $z_0 = \exp(x_0)$ with 47+ bits in average, where 47=100-53 is the extra accuracy of twofold over standard `double`.

Extra accuracy in $L_1$ norm for $\log(y)$ and is 45=98-53 bits. Extra accuracy in $L_0$ norm is 40=93-53 bits for $\log(y)$ and 42=95-53 bits for other functions.

For `float` type, extra accuracy in $L_1$ norm is 18=42-24 bits for all functions, and in $L_0$ norm is 12=36-24 bits for $\log(y)$ and 14=38-24 bits for other functions.

Is such extra accuracy enough? I think "error" part of twofold might be even less accurate, though more extra bits is better of course. Other authors like Masotti [4], propose "error" part mantissa be 2/3 of the main "value" part; that is $z_1$ to hold 35-36 bits for `double` and 16 bits for `float`.

Twofolds of `double` can meet this criterion with both $L_0$ and $L_1$ metrics. Twofolds of `float` can meet it only with $L_1$ but not with $L_0$ metric.

Anyway, I think proving consistency of twofolds for tracking math accuracy is not subject for deductive analysis. More important is confirming by practice, experimentation with real-world applications. Only wide practice can show if verifying accuracy of math computations with twofolds is worth investments.

## Conclusion

I seek for an easy way for automatic control of rounding errors, for simplifying programming of math.

Twofold function, like $z_0 + z_1 = \text{texp}(x_0 + x_1)$, bitwise reproduces standard C/C++ math library result $z_0 = \exp(x_0)$, so that $z_1$ assess deviation $\Delta z_0 = z - z_0$ from exact value $z = e^{x_0+x_1}$. Ideally, $z_1$ should equal value of $\Delta z_0$ correctly rounded to nearest-even floating-point number. In reality, we have to seek for a balance of accuracy versus performance.

This article proposes the specific way for the balancing, with $z_1$ average accuracy like 47+ significant bits if `double` format (45+ bits for $\log(y_0)$ function), and performance 10-20x times higher than quad-math library as supported with GNU compiler. I think such accuracy and performance might be good enough for regularly tracking rounding errors in majority of standard-precision calculations.

I think one cannot mathematically deduce if twofolds are “good enough”. Proving that is rather subject for experimenting with variety of practical applications. However, let me express my point, why I think that verifying mathematic computations with twofolds must be technically consistent and useful.

Technically, assessing accuracy is fundamentally easier than improving it. So generally, $z_1$ does not need be very strict. Enough if accuracy of twofold function is not worse than original standard function.

That is, if we consider $z_0 + z_1 = \text{texp}(x_0 + x_1)$ like approximation for $z = e^x$, where $x = x_0 + x_1$, then twofold $z_0 + z_1$ must not deviate from exact $z$ by more than $z_0 = \exp(x_0)$ deviates, if the input $x_0 + x_1$ approximates $x$ with same or better accuracy than $x_0$ alone.

Simple but useless way to grant this “not worse than standard” property is let $z_1$ be always zero.

Being not worse than standard is the key property for automatic testing; it guarantees twofolds would never raise a panic in vain, if not sure. In worst case, twofolds may underestimate accumulated errors, and fail catching accuracy problems with standard-precision code.

More accurate $z_1$ increases chances that testing with twofolds is profitable, can adequately measure inaccuracy and catch majority of accuracy problems.

High performance on modern processors must allow checking on fly, in parallel with main computations. Future processors can reduce cost of twofolds even more. Ultimately, I would propose twofolds like new kind of floating-point numbers, with built-in control of rounding errors.

Consider twofolds as sort of “managed runtime” for floating-point computations. Computers penetrate everywhere, so need mass of programmers working faster, and frankly getting less skilled in average. A managed runtime can mitigate cost of programming; allow coding easier with higher quality.

Despite twofolds cost, balance to benefits looks promising, as people productivity is more important.

# Twofolds update

## Older CPUs

As people started asking me about twofold arithmetic, I realized that many might still have older CPU versions that do not support fast fused-multiply-add (FMA) instructions. Thus, I decided to support old processors, and implemented alternative algorithms that do not depend on fast-FMA. This alternative code is somewhat slower, but I think it is fast enough to give you perception of twofolds.

Specifically, I have implemented older (pre-FMA) algorithm for exact multiplication, usually credited to Dekker and Veltkamp. Then I use Dekker-Veltkamp multiplication for simulating $\mathrm{fma}(x, y, z)$ function in important special case if $xy \approx -z$. Twofolds use FMA in this special case for taking exact remainders in square root and dividing functions, like $r = \mathrm{fma}(-q, b, a)$ where $q \approx a/b$.

Following are the formulas, as I borrow them from Shewchuk paper [9].

First algorithm splits a floating-point number $x$ into "higher" and "lower" halves $h$ and $l$, each holding around half of significant bits. For standard `double`, $h$ and $l$ each would hold 26 bits of $x$, so 52 bits in overall, and remaining 53'rd bit of $x$ is encoded with sign of $l$.

Let constant $p$ be amount of bits in mantissa, $p = 53$ for `double` and $p = 24$ for `float` type:

<u>Algorithm DV1.1 (Dekker)</u>: $h + l = \mathrm{split}(x)$

(1) Integer $s = \lceil p/2 \rceil$ -- round upward
(2) Floating $a = (2^s + 1) \cdot x$ -- round to nearest-even
(3) Floating $b = a - x$ -- round to nearest-even
(4) Floating $h = a - b$ -- exact by Sterbenz lemma
(5) Floating $l = x - h$ -- exact by Sterbenz lemma

Second algorithm multiplies two floating-point numbers $x$ and $y$ by parts. Result is coupled-precision number $z_0 + z_1$ with value exactly equal to $xy$, unless an overflow or underflow occurs:

<u>Algorithm DV1.2 (Veltkamp)</u>: $z_0 + z_1 = x \cdot y$

(1) Dotted $z_0 = x \cdot y$ -- with 1x-precision, round to nearest-even
(2) Coupled $x_0 + x_1 = \mathrm{split}(x)$ -- by Dekker algorithm
(3) Coupled $y_0 + y_1 = \mathrm{split}(y)$ -- by Dekker algorithm
(4) Dotted $e_0 = z_0 - (x_0 \cdot y_0)$ -- exact
(5) Dotted $e_1 = e_0 - (x_0 \cdot y_1)$ -- exact
(6) Dotted $e_2 = e_1 - (x_1 \cdot y_0)$ -- exact
(7) Dotted $z_1 = (x_1 \cdot y_1) - e_2$ -- exact

Following is algorithm for simulating FMA in special case if $xy \approx -z$. In such conditions, Sterbenz lemma guarantees that intermediate result is exact, so final result is rounded only once like expected for FMA:

<u>Algorithm DV2</u>. $r = \mathrm{fma}(x, y, z)$, if signs of $xy$ and $z$ opposite, and $|z|/2 \le |xy| \le 2|z|$

(1) Transform $p + t = x \cdot y$ -- exact transform via Dekker-Veltkamp algorithm (DV1)
(2) Let $r = (z + p) + t$ -- result of $z + p$ is exact by Sterbenz lemma

In overall, Dekker-Veltkamp exact multiplication includes 7 multiply and 10 add/subtract operations, so 17 processor instructions in overall. This is 8.5 (=17/2) times more than the algorithm using FMA.

However, performance gap must be much fewer in terms of processor ticks. FMA-based algorithm takes at least 3 ticks on processor like Intel Haswell, as FMA itself takes 2. In turn, processor able to multiply in parallel with add/subtracts. So critical path for Dekker-Veltkamp is 10 ticks for add/subtract operations.

Thus expected performance gap must be “only” around 10:3 on processor like Intel Haswell.

According to my testing, the gap is actually even fewer, about 2.15 times for twofold multiplying and 10-40% for dividing and square root. See my results for GNU compiler, plain C, Haswell 2.25 GHz, manually vectored for AVX, performance per one CPU core, measured by million operations per second (MOPS).

Table: Twofold arithmetic performance gap, with and w/o FMA

| | | tadd | tmul | tdiv | tsqrt |
|---|---|---|---|---|---|
| | double | 1145.6 | 1480.49 | 162.928 | 108.706 |
| With FMA | float | 2291.85 | 2957.52 | 645.312 | 432.816 |
| | double | 1145.82 | 687.599 | 146.512 | 108.54 |
| W/o FMA | float | 2290.19 | 1379.59 | 466.543 | 401.155 |

Good question is why would not we rely on standard $\mathrm{fma}(x, y, z)$ function from C/C++ math library?

In absence of fast-FMA hardware support, GNU compiler sometimes replaces $\mathrm{fma}(x, y, z)$ with simply $xy + z$. Such replacement makes code fast, but completely damages FMA-based algorithm for taking remainders and for exact multiplication, which explore tricks like $\mathrm{fma}(x, y, -xy)$.

With Microsoft compiler, $\mathrm{fma}(x, y, z)$ is “honest” but too slow, around 1000 (thousand!) times slower than hardware according to my testing. Certainly, Dekker-Veltkamp would be very much faster.

Besides, my twofolds code via Dekker-Veltkamp includes vectoring for SSE2 and AVX.

For switching twofolds code between older and newer processors, see Compiler flags section below.

## Compiler flags

Twofolds code default behavior changed to support older processors.

In previous versions, twofold arithmetic assumed processor supports Intel AVX with fast-FMA. Code version that I provide with this article supports older processors as well. Here “older” means any one with IEEE-754 compatible `float` and `double` types. Any processor like Intel Pentium 4 or newer fits.

Default code supports only scalar calculations, and uses Dekker-Veltkamp algorithms for multiplying and taking remainders. If your processor supports SSE version 2, twofold can leverage of this by vectoring for SSE2. For such vectoring, please compile with –DSSE2 option.

Some processors support AVX but not fast-FMA. Compiling with –DAVX option unlocks vectoring for AVX, though code still uses Dekker-Veltkamp algorithms so do not depend on fast-FMA.

If processor supports both AVX and FMA, you may compile with –DFMA option to unlock the twofolds code branch that supplies the best performance.

Please do not forget to tell compiler what CPU architecture you target. For GNU compiler, use option like –msse2, –mavx, or –mfma. If Microsoft, you may target /arch:SSE2, /arch:AVX, or /arch:AVX2.

Finally, twofolds code heavily depends on rounding tricks with expressions like $(x + y) - y$. If compiler optimizes-out such expression into just $x$, this would completely damage twofold arithmetic. Please tell compiler not to optimize such expressions. Compile with /fp:strict or –frounding-math option.

So use –O3 optimization level on GNU, not –Ofast. For Microsoft, use /Ox optimization.

Following table summarizes compiler flags that I recommend for best performance:

| Processor | GNU compilers | Microsoft |
|---|---|---|
| Any w/IEEE types | –O3 –frounding-math | /Ox /fp:strict |
| Vectoring for SSE2 | –O3 –frounding-math –msse2 –DSSE2 | /Ox /fp:strict /arch:SSE2 /DSSE2 |
| AVX, but w/o FMA | –O3 –frounding-math –mavx –DAVX | /Ox /fp:strict /arch:AVX /DAVX |
| AVX and fast-FMA | –O3 –frounding-math –mfma –DFMA | /Ox /fp:strict /arch:AVX2 /DFMA |

Note that /arch:SSE2 is only available with 32-bits variant of Microsoft compiler for x86. Compiling for 64-bits, you may just omit /arch option, as Microsoft compiler assumes SSE2 for all x86-64 processors.

## Compiler bugs

Unfortunately, the bug in GNU gcc/g++ compilers that I use (Cygwin, GNU gcc/g++ 4.8.3 and 4.9.2), may severely affect my style of programming with Intel AVX intrinsics. For bug details please follow the link:

http://sourceware.org/ml/cygwin/2014-07/msg00056.html

Compiler may sometimes (not always) misalign AVX vectored data of __m256d/__m256 type in stack. As result, functions returning AVX vectored result may crash with “Segmentation fault”, like for example:

```
static __m256d dset1x4 (double x) { return _mm256_set1_pd(x); }
```

Such crash happens intermittently, that is occasionally, depending on some random circumstances. This is very good piece of luck why majority of AVX tests passed for me with GNU gcc/g++. Of course I cannot recommend you relying on such luck.

If testing crashes for you, I recommend trying another compiler, preferably gcc/g++ with the bug fixed. Anyway, you may simply omit testing of vectored variant of twofold arithmetic. Twofold exponent and logarithm functions do not support vectoring, so this compiler bug does not affect them.

GNU gcc/g++ vectoring for SSE2 is not impacted. And Microsoft compiler is not impacted at all.

Sorry for this limitation! I will try to propose a better workaround for this problem in my next article.

## Twofold bugs

In my previous article entitled “Twofold fast arithmetic” [1], sign is mistakenly messed in the following twofold subtraction formulas. This mistake would cause inaccurate results of twofold subtraction.

Correct variant:

Algorithm 2.2: $a - b \to d + t$ for arbitrary $a$ and $b$

(1) $d = a \ominus b$
(2) $b' = d \ominus a$
(3) $a' = d \ominus b'$
(4) $b^{\#} = b \oplus b'$
(5) $a^{\#} = a \ominus a'$
(6) $t = a^{\#} \ominus b^{\#}$

Corresponding bug also affects C/C++ experimental code that I provide with earlier articles [1] and [2]. So for trying twofold software better download its fresher version that I provide with this new article.

# How to download

Text and code available at this project web site: https://sites.google.com/site/yevgenylatkin/

Please do not hesitate to contact me if any comments and/or questions: yevgeny.latkin@gmail.com

# Acknowledgements

I would like to thank following people with whom I have discussed twofold arithmetic ideas:

- Nikita Astafiev (Intel)
- Dmitry Baksheev (Intel)
- Marius Cornea (Intel)
- Bob Hanek (Intel)
- Victor Kostin (Intel)
- Evgeny Petrov (Intel)
- Ivan Golosov (UniPro)
- Alexander Semenov (UniPro)

Company web sites FYI:

- Intel Corporation: http://intel.com
- UniPro: http://unipro.ru/eng/index.html